# Compound cryopump for fusion reactors

M. Kovari*, R. Clarke, T. Shephard

*EURATOM/CCFE Fusion Association, Culham Science Centre, Abingdon, Oxon, OX14 3DB, UK*
*Corresponding author. Tel.: +44 (0)1235-46-6427. E-mail address: michael.kovari@ccfe.ac.uk

ABSTRACT

We reconsider an old idea: a three-stage compound cryopump for use in fusion reactors such as DEMO. The helium "ash" is adsorbed on a 4.5 K charcoal-coated surface, while deuterium and tritium are adsorbed at 15-22 K on a second charcoal-coated surface. The helium is released by raising the first surface to ~30 K. In a separate regeneration step, deuterium and tritium are released at ~110 K. In this way, the helium can be pre-separated from other species. In the simplest design, all three stages are in the same vessel, with a single valve to close the pump off from the tokamak during regeneration. In an alternative design, the three stages are in separate vessels, connected by valves, allowing the stages to regenerate without interfering with each other. The inclusion of the intermediate stage would not affect the overall pumping speed significantly.

The downstream exhaust processing system could be scaled down, as much of the deuterium and tritium could be returned directly to the reactor. This could reduce the required tritium reserve by almost 90%.

We used a well-established free Direct Simulation Monte Carlo (DSMC) code, DS2V. At very high upstream densities (~$10^{20}$ molecules/m$^3$ and above) the flow into the pump is choked. Enlarging the aperture is the only way to increase the pumping speed at high densities. Ninety percent of the deuterium and tritium is successfully trapped at 15 K (assuming that the sticking coefficient is 80-100% on the 15-22 K surface). On the other hand, the remaining 10% still exceeds the small amount of helium in the gas input.

## 1. Introduction

A magnetic confinement fusion reactor needs a set of pumps near the divertor to capture helium ash, which is mixed in with deuterium and tritium. While the helium content of the core plasma may be of the order of 10%, the scrape-off layer which reaches the divertor may be ~2% helium [1]. Consequently each helium atom reaching the pump is accompanied by 25 atoms of tritium. The fractional burn-up of tritium in this case is 4% (defined as fusion reactions / tritium atoms supplied), providing that no helium was injected into the reactor. (DD reactions are ignored.) The helium concentration may be only 1% or less if a "puff and pump" approach is used, in which excess fuel is deliberately added to the scrape-off layer to prevent contamination from impurity seeding from reaching the core [2]. It may be possible to achieve a tritium burn-up as high as 4% by injecting tritium only in the form of DT pellets, while fuelling the edge plasma with additional deuterium [3].

In ITER the plasma exhaust will be pumped using large cryopumps near the divertor region, although it has been suggested that other pumping technologies may be more suitable for a power plant [4]. Here we assume that a cryopump is used. After the cryopump is regenerated by raising its temperature, the gas released must be processed to extract the tritium and deuterium for return to the reactor. There are several ways in which the quantity of gas requiring processing can be reduced, as follows.

(a) Simplest of all is to return some of the pumped gas directly to the fuelling system, while processing the remainder. In this case the fuel injected would be contaminated with helium. This is acceptable if the helium density gradient from core to divertor is still sufficient. Any scheme such as this in which gas is returned directly without leaving the vicinity of the plasma is known as Direct Internal Recycling (DIR) [5].

(b) Pre-separation of helium has been proposed for ITER [6]. In this scheme all gases would be captured together in the main pump. The gas mixture released during regeneration flows to a secondary pump, which consists of three chambers. In the first chamber the impurities are condensed at 20 K. In the second chamber hydrogen isotopes are condensed at 5 K. In the third chamber helium is adsorbed on charcoal at 5 K, or alternatively pumped out by a mechanical system. 90% of the hydrogen isotopes are recycled directly into the torus.

(c) It is possible to install additional, physically separate pumps to capture gases other than helium, enriching the remaining divertor gas in helium.

(d) Another form of pre-separation is to feed the gas mixture after regeneration into a pellet machine. This would freeze the D and T, releasing the helium for processing. The helium concentration in the extruder itself must be very low to ensure the strength of the pellet [7], but it may be possible to design the gas condensing stage to allow the helium to escape – effectively turning the pellet machine into a miniature cryo-distillation device.

(e) In this paper we consider a cryopump that separates the gases at the point of initial capture: a compound cryopump. The first stage is a radiation shield at 80 K. The second is a set of charcoal-coated plates at 15-22 K, pumping deuterium and tritium. The third stage is a set of charcoal-coated plates at ~4.5 K, pumping helium. The helium is released for processing by raising the 4.5 K surface to ~30 K. Then deuterium and tritium are released at ~110 K. Provided the helium content in this fraction is low enough, it could be allowed to re-enter the reactor directly, or pumped out to feed the pellet injector, without any processing at all. This may reduce the tritium reserve required. (This idea was proposed for ITER [7, 8], but was never adopted.)

Options (c) and (e) can be used together, with some 2-stage pumps at 15-22 K pumping D and T only, and a few compound pumps collecting helium as well. The 15-22 K stages could be cooled by a hydrogen liquefier (operating indirectly via a gaseous helium coolant). This would reduce helium inventory and the power required for cryogenic refrigeration.

The compound cryopump was first invented to protect the low temperature stage from large loads of condensable gases such as oxygen, nitrogen and water [9], and commercial cryopumps use a compound design, with a condensation stage doubling as a radiation shield at 40 to 70 K, and an adsorption stage at 11 to 20 K. Compound cryopumps have also been trialled for use in fusion reactors [10, 11].

## 2. Direct Simulation Monte Carlo and the DS2 and DS2V codes

To predict the behaviour of a compound cryopump, a Monte Carlo approach is both simple and accurate, and capable of simulating any geometry. To ensure that flow in the transition regime can be represented, the Direct Simulation Monte Carlo (DSMC) technique was used, as it takes account of intermolecular collisions. DSMC has already been used to model cryopumps – for example, [12] simulates the ITER torus cryopump at ½ scale, using a combination of Test Particle Monte Carlo (TPMC) and DSMC techniques.

Calculations were done using a well-established free DSMC code, DS2V, which incorporates DS2 [13] and a visual interface. We made small modifications to DS2 to extend its output, not changing the underlying algorithm. The modified code can be made available on request, providing the permission of the originator of the code, Professor Bird, is obtained, enabling other workers to replicate or extend our results. A parallel version of DS2 is available [14], but was not used in this work.

The principle of DSMC is the same as for TPMC, in that one model particle represents a large number of real molecules. In TPMC it is convenient to track particles one at a time from when they are created to when they escape from the model, so there is no global time variable, and a given solution is applicable irrespective of the rate of gas throughput. In DSMC all the particles are tracked simultaneously, with time being a variable. Collisions between particles are simulated explicitly. However, the number of model particles is large, so it would not be efficient to consider collisions between all possible pairs. In the Bird method used by DS2 [15] a certain number of pairs are selected in each cell at each time step, and the collision probability is adjusted appropriately. DS2 can model steady or unsteady plane or axially symmetric flows. Energy exchange between translation and rotation is included. Molecular vibration was neglected in this work, as small molecules are usually in the vibrational ground state at room temperature and below. The criteria for a good DSMC calculation are that, at every location in the flow, the time step should be small in comparison with the mean collision time and the cell size should be small in comparison with the mean free path. The time step varies across the flowfield, being automatically adapted to a specified fraction of the local mean collision time. The default value of this fraction is one third.

Axially symmetric flow was used in this work. The molecules are moved in three dimensions and the collisions are three-dimensional events. However, at the end of each molecule move, molecules are moved to the zero azimuth angle. Optional radial weighting factors enable a uniform distribution of molecules over the computational plane. To achieve this some molecules are discarded when they move closer to the axis and some are duplicated when they move away from it.

At material surfaces, we have assumed diffuse reflection with complete translational and rotational accommodation to the surface temperature, as recommended by Bird [15]. A fraction of the molecules striking a surface may be specified as being adsorbed at the surface. A surface with 100% adsorption is equivalent to a vacuum interface, and a fraction less than one was used to model a cryogenic surface that traps molecules.

In the benchmarking below, collisions were simulated with the hard sphere model, as used by the authors with whose results this work was compared [16]. In the cryopump simulations, the variable hard sphere (VHS) model was used, where the molecular diameter varies with collision energy so that the viscosity varies as a power of the temperature. The exponent recommended for hydrogen is 0.67 [15]. (The hard sphere model is a special case of VHS where the diameter is fixed and the exponent is 0.5.)

## 3. Benchmarking the code

The DS2 code was validated against known benchmarks [16] for flow through a tube. The results are in Figure 1.

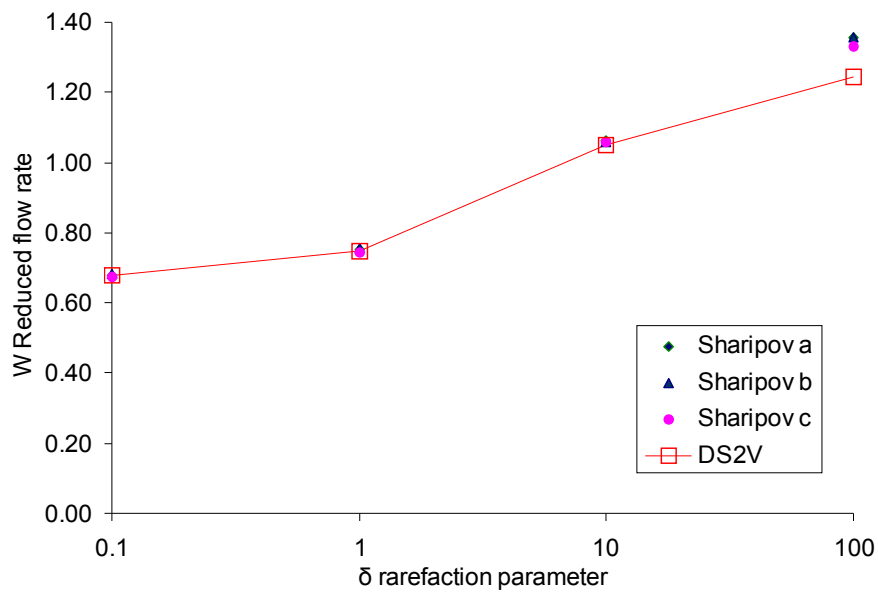

**Figure 1**. Benchmark results. The labels a, b and c refer to three different simulations compared by [16]. W is the ratio of the mass flow rate through the tube to the theoretical flow rate through an aperture in the free molecular regime. The quantity $\delta$ [16] is inversely proportional to the Knudsen number.

The case with densest gas ($\delta = 100$) differs from the benchmark by 8% - more model particles would be required for this regime. Simulations were also done to make sure the program successfully calculated the behaviour of gas mixtures, using two gases with molecular masses equal to 1 and 1.5 arbitrary units. As expected, the pumping speed for each gas was the same whether or not the other gas is present, and inversely proportional to the molecular mass. Pumping speed S in this report is defined as

$$S = \frac{R}{n_0},$$

where $R$ = rate of adsorption of molecules,
and $n_0$ = upstream number density (specified at the flow entry surface).

The benchmarking indicates that DS2 and DS2V can be used with about 1% accuracy, provided the ratio of mean collision separation to mean free path is much less than one.

## 5. Compound cryopump – single chamber

Figure 2 shows the DS2V model of a single-chamber compound pump. Table 2 gives the temperatures and sticking coefficients. The sticking coefficients and other parameters are approximately appropriate if charcoal is used as the cryosorbent.

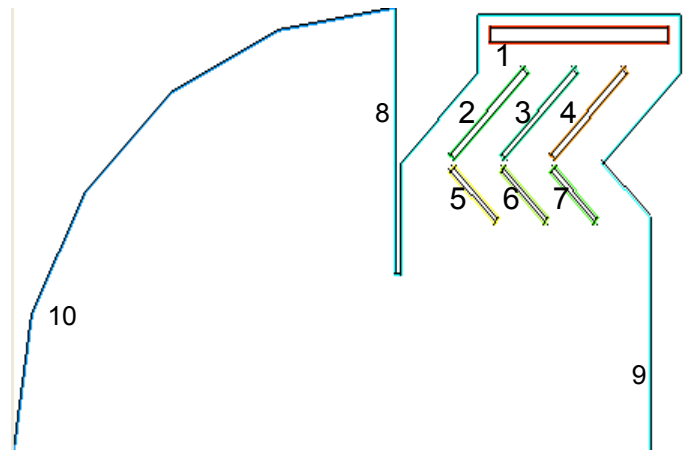

**Figure 2**. Half cross-section of single-chamber 3-stage pump, with inlet reservoir on the left. The total axial length of the model is 1.35 m, and the overall radius is 0.76 m. The aperture has radius 0.3 m.

**Table 1**

| Surface (Figure 2) | Description | Temperature K | Sticking coefficient $D_2$ | $T_2$ | He |
|---|---|---|---|---|---|
| 1 | Stage 3 | 4.5 | 0.9 | 1 | 0.3 |
| 2, 3, 4 | Stage 2 | 15 | 0.8 | 1 | 0 |
| 5, 6, 7 | Stage 1 | 80 | | | |
| 8 | Outside wall | 273 | | | |
| 8 | Inside walls | 80K | | | |
| 9 | Inlet | (Gas at 273 K) | | | |

Preliminary runs showed that for this model, radial weighting factors have very little effect on convergence time, but, unlike in the case of the short tube, convergence is faster if the flow field at zero time is already populated with the full upstream density.

As before, the bigger the inlet reservoir, the more accurate the assumption that the gas is at rest at the inlet surface. Figure 3 shows that a dome of radius 0.75 m (2.5 times the aperture radius) gives results within a few percent of the converged value. This radius has been used for all the calculations described below. A larger dome requires many more molecules and therefore a much longer computing time. While this dome is much smaller than that used in the benchmarking above, we are dealing here with concept models and design principles, so the same level of accuracy is not required.

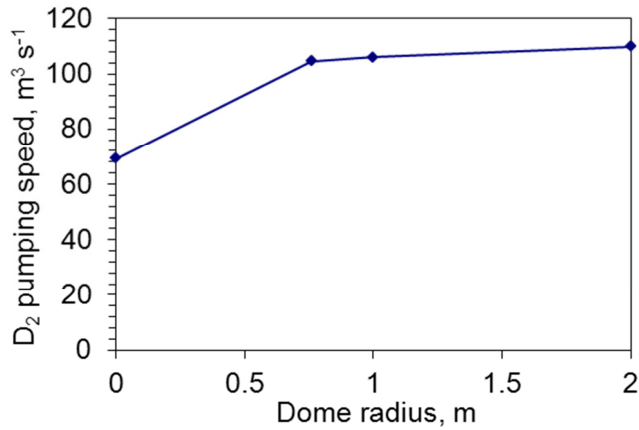

**Figure 3.** Pumping speed for $D_2$ *vs* radius of inlet reservoir (dome)

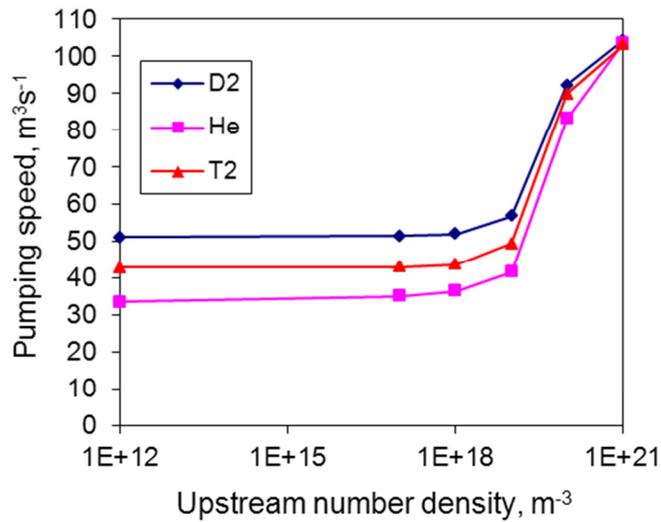

**Figure 4**. Pumping speeds for single-chamber 3-stage pump

As expected, the pumping speed is constant in the molecular regime and increases through the transition regime (Figure 4). The increase then slows at very high densities because the flow becomes choked (transonic). Qualitatively, results look similar to published results [17]. At very high density, the gas behaves as a single fluid, giving a single pumping speed for all species.

The flow reaches Mach 1 in the vicinity of the aperture, as expected for choked flow, becomes supersonic downstream of the aperture, but then passes through a compression shock. The choked flow through the orifice limits the pumping speed [18].

## 7. Separation of gases

It is important to know how much of each gas is on stage 2 compared to stage 3, and the composition of the gas adsorbed on each surface. For example, Table 3 and Table 4 gives results for the effectiveness of the pump in separating gases in the molecular regime (gas composition is 49% $D_2$, 49% $T_2$, 2% He, n = $10^{12}$ m$^{-3}$).

**Table 2.** Location where each gas is adsorbed

|  | Stage 2 | Stage 3 | Stage 2 + 3 |
|---|---|---|---|
| $D_2$ | 88.8% | 11.2% | 100% |
| He | 0.00% | 100.00% | 100% |
| $T_2$ | 90.9% | 9.1% | 100% |

**Table 3.** Composition of gas on each panel

|  | Stage 2 | Stage 3 |
|---|---|---|
| $D_2$ | 53.9% | 52.3% |
| He | 0.00% | 12.4% |
| $T_2$ | 46.1% | 35.2% |
| Sum | 100% | 100% |

For example, only 9% and 11% of $T_2$ and $D_2$ respectively are trapped on stage 3, but because they each make up 48% of the upstream gas, they constitute 52% and 35% of the gas adsorbed on stage 3. Figure 7 shows that about 90% of the $D_2$ and $T_2$ are trapped on stage 2, and this fraction increases to about 95% for very high densities.

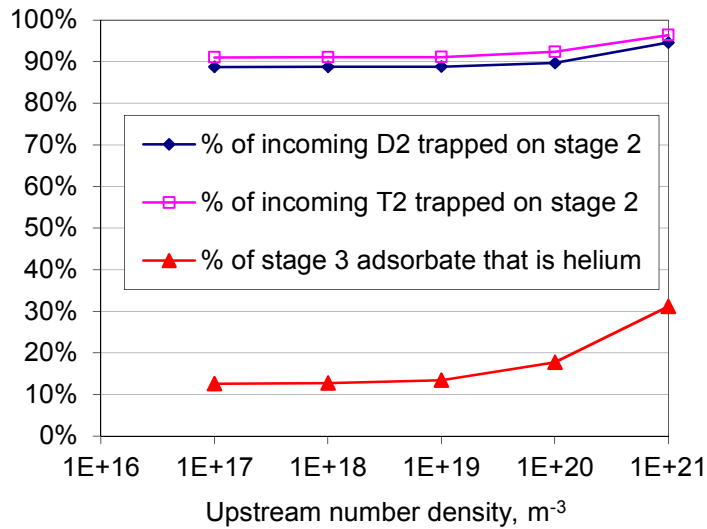

**Figure 5**. Indication of where each gas is trapped.

It is possible that helium will be co-adsorbed onto stage 2. Figure 8 shows that if helium has a sticking coefficient of 1%, the gas adsorbed in stage 2 would be about 0.2% helium – ten times less than in the incoming gas.

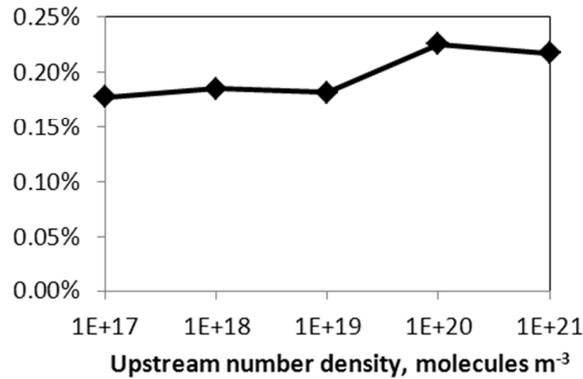

**Figure 6**. Percentage of stage 2 adsorbate that is helium. Sticking coefficient is 0.01, and incoming gas is 2% helium.

## 8. Obstruction by Stage 2 panels

To compare the performance to a similar two-stage pump the stage 2 panels were removed. In the molecular regime, deuterium is pumped 7.8% faster with the stage 2 panels present, and the helium pumping speed is reduced by only 15% by the stage 2 panels. At the highest density, omitting the panels has only about 1% effect for each species.

## 9. Temperature of the stage 2 panels

Figure 9 shows results with reduced sticking coefficients on the stage 2 panels, as would occur if the panel temperature were higher. The effect on the pumping speed for $D_2$ is small, but there is a substantial effect on where the gases are adsorbed – significantly more hydrogenic species are pumped on the 4.5 K panel.

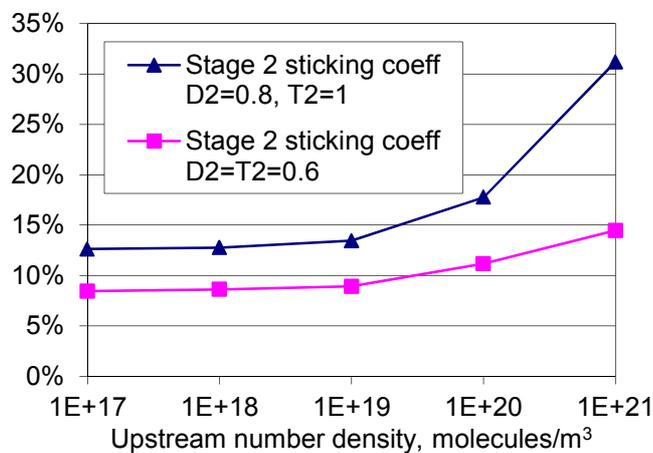

**Figure 7.** Percentage of the gas trapped on stage 3 that is helium, different stage 2 sticking coefficients

## 10. Long cryopump

Figure 10 shows a longer pump with 10 sets of panels instead of three. The larger number of panels will increase the capacity, and may increase the pumping speed. The other dimensions are the same. The streamlines show a conical vortex on the back wall, and a small vortex downstream of the edge of the aperture.

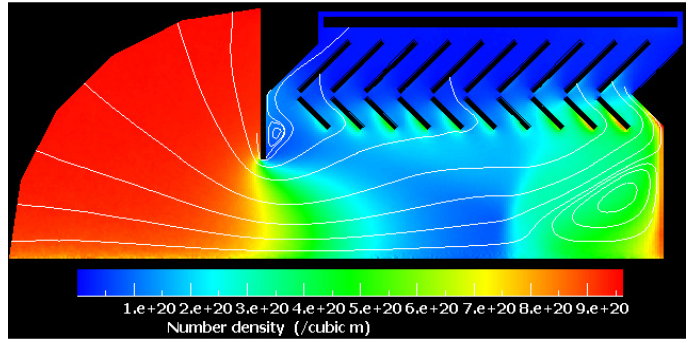

**Figure 8**. Single-chamber 3-stage pump with 10 sets of panels. The white lines are streamlines.

The pumping speeds of the long pump were evaluated in comparison with the short pump. When the upstream density is high, there is no difference at all, because the flow is choked at the inlet. The biggest difference is for helium at low density, but even then the long pump gives only 52% more pumping speed, even though it has 3⅓ times as many panels. It is clear that the aperture of the pump is much more important than the pumping area.

## 11. Fuel cycle

Using a standard 2-stage cryopump with no pre-separation, the helium must be removed before the accompanying tritium can be returned to the reactor. If the reactor must be able to run for a period of time when the processing plant is out of action, then it requires a tritium reserve proportional to the gas capture rate necessary for helium extraction. For given fusion power, this reserve is inversely proportional to the helium content of the exhaust gas. The calculated reserve can easily reach undesirable and unachievable levels of hundreds of kg [19]. The compound pump allows most of the fuel mixture to be returned directly to the reactor, eliminating much of this tritium reserve requirement. If only 10% of the $D_2/T_2$ captured is trapped on stage 3 (Table 5) then the flow of gas through the processing plant will be reduced to about 10% of the value for a 2-stage pump with no pre-separation. One can formalise this in Table 5 and equations 5 - 7.

**Table 4**. Symbols and assumptions

| | | |
|---|---|---|
| $R_f$ | Rate of fusion reactions (Fusion power=3000 MW) | $1.1 \times 10^{21}$ s$^{-1}$ |
| $f_b$ | Fractional burn-up of tritium | 4% |
| $f_3$ | fraction of captured hydrogen isotopes trapped on stage 3 (Table ) | 10% |
| $t_r$ | Time for which reactor must run while all processing plant is out of action | 1 day |
| | Number of pumps | 16 |
| | Length of pumping/regeneration cycle | 1200 s |
| $m_T$ | Mass of tritium atom | 5.0 10$^{-27}$ kg |

The flow rate of all atoms (T, D and He) through the plasma exhaust processing plant is given by

$$R_f\left(1+\frac{1-f_b}{f_b}2f_3\right) \approx 6R_f. \tag{5}$$

By comparison, for a 2-stage pump with no pre-separation it is given by

$$R_f \frac{2-f_b}{f_b} = 49R_f, \tag{6}$$

a factor of 8 more. The tritium reserve required for the two cases is

$$R_f t_r m_T \left(1 + f_3\left(\frac{1}{f_b} - 1\right)\right), \text{ or } R_f t_r m_T \frac{1}{f_b}. \tag{7}$$

The use of the compound pump reduces the tritium reserve from 12 kg to 1.6 kg. If the separation is effective enough ($f_3/f_b \ll 1$), the tritium reserve reduces to the theoretical minimum, $R_f t_r m_T$, even if the burn-up fraction is small.

Figure 11 illustrates the gas system. The tritium breeding system is not shown, but its output of tritium will be much less than the rate of tritium flow through the fuelling system.

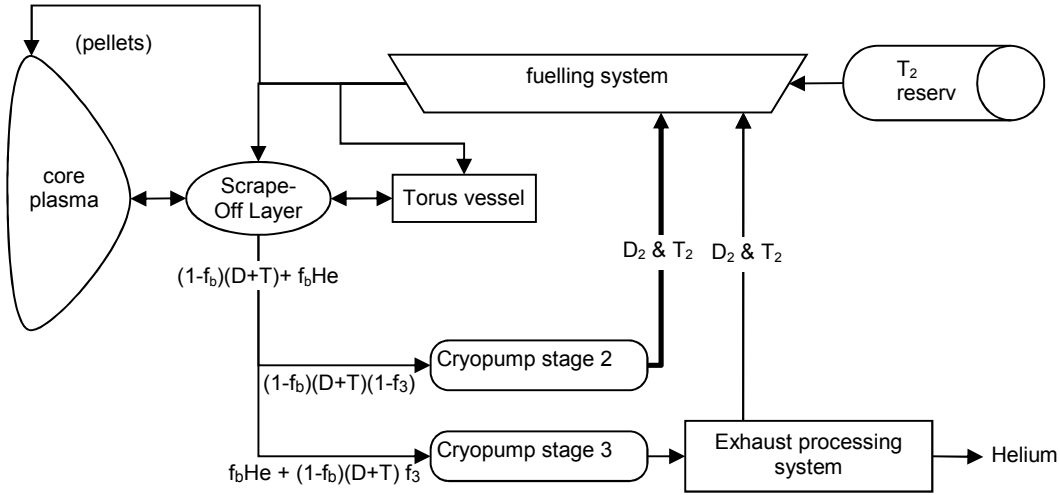

**Figure** 9. Gas system with 3-stage pump. The numbers of atoms of each species being trapped on the cryopump are shown, per tritium atom injected by the fuelling system.

## 12. Regeneration

For the single-chamber compound pump to work, it must be possible to release the helium into the pump by raising stage 3 to 30 K without causing too much temperature rise in stage 2. Figure 12 illustrates the envisaged layout.

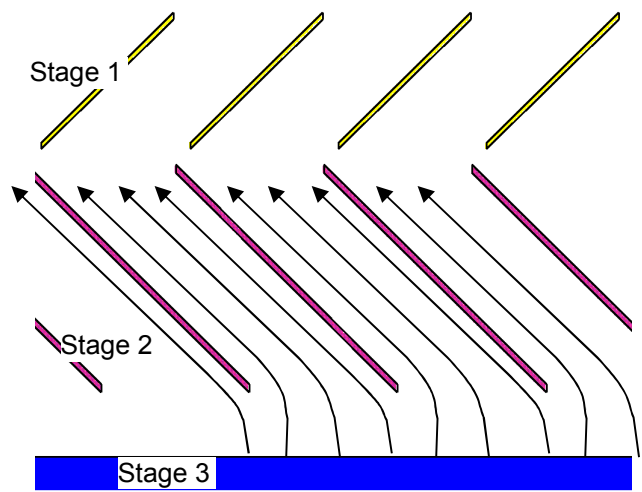

**Figure** 10. Illustration of bulk gas flow from regenerating stage 3.

The resulting flow rate of released helium is given by the mass of gas and the time over which the pump is re-generated. With parameters from Table 5, 3 g of gas is released over an area of 5 m$^2$. The Reynolds number is given by

$$Re = \frac{\rho v L}{\eta}, \tag{3}$$

which for a mass m of gas coming out of a surface of area A during time t reduces to

$$Re = \frac{m}{At} \frac{L}{\eta}. \tag{4}$$

If we take the duration of outgassing as 75 s, and the length scale as 10 cm, the Reynolds number is only 0.5, which is far too small for turbulent effects to be important.

Given these conditions, the flow of heat to the second stage will consist primarily of two effects: (a) the direct convection of heat carried by the released helium (assuming it is all thermalised on contact with the second stage); and (b) thermal conduction between the stages through the body of the gas. (Radiation is of course always present.) A simple estimate suggests that the temperature of stage 2 will increase by ~0.04 K due to (a), and ~6 K due to (b) (assuming that the regeneration causes a pressure excursion that lasts for 200 s). In practice the refrigeration will be adequate to maintain a constant temperature, and in any case stage 2 will not approach 50 K, at which temperature desorption of $D_2$ and $T_2$ starts to become significant. These estimates assume that the $D_2$ and $T_2$ adsorbed in stage 3 (which greatly exceed the quantity of helium) are not desorbed at 30 K. More careful analysis of the regeneration process could be carried out using DS2V, which calculates the energy transferred between gas molecules and surfaces.

## 13. Discussion and Conclusions

A compound cryopump concept for fusion reactors has been modelled and it is shown that some pre-separation of hydrogenic species from helium is possible without adversely affecting capture efficiencies. This would allow Direct Internal Recycling of fuel, and reduce the amount of isotopic separation that needs to be done by an order of magnitude, simplifying the fuel cycle both for DEMO and for fusion power plants.

The simplistic assumption was used that the sticking coefficient of each gas is independent of pressure and gas loading, which is certainly not the case. It is also possible that helium would be co-adsorbed by deuterium and tritium onto the charcoal. Experimental work is required to investigate these factors.

Benchmarking indicated that DS2V can be used with sufficient accuracy. Checks also suggested that gas mixtures can be successfully simulated, and that valid results can be obtained even when shocks are present. The input file for a typical simulation is given in Supplementary Data. The same information in human-readable form, as output by the code, is also given. The input file will run successfully under DS2V without modification, but code changes are needed to obtain all the required species-specific output.

An alternative pump design would have three chambers, for pumping helium, hydrogenic species, and impurities respectively. The chambers would be separated by discs that can act as valves, allowing the stages to regenerate without interfering with each other.

Detailed results were obtained for the single-chamber pump. When the gas entering the pump is at very high density (~$10^{20}$ molecules/m$^3$ and above) the flow into the pump in its current design is choked. Enlarging the aperture is the only way to increase the pumping speed at high densities, and even at low densities it is the most important way. Only limited separation of helium is achieved. Ninety percent of the deuterium and tritium are successfully trapped at 15 K, but the remaining 10% still exceeds the small amount of helium on the 4.5 K panels. The specificity achieved will depend on the temperature of the intermediate panels and the resulting sticking coefficients. The 15 K panels do not greatly reduce the overall pumping speed. Even at low densities the helium pumping speed is only reduced by 15% by the presence of these panels.

**Acknowledgments**

We would like to thank Professor G.A. Bird for providing the DS2V code, and Stylianos Varoutis, Chris Day, David Ward and Elizabeth Surrey for helpful advice. This work was funded by the RCUK Energy Programme under grant EP/I501045 and the European Communities under the contract of Association between EURATOM and CCFE. The views and opinions expressed herein do not necessarily reflect those of the European Commission. This work was carried out within the framework of the European Fusion Development Agreement.

**Figure Captions**

**Figure 1**. Benchmark results

**Figure 2**. Half cross-section of single-chamber 3-stage pump, with inlet reservoir on the left. The total axial length of the model is 1.35 m, and the overall radius is 0.76 m. The aperture has radius 0.3 m.

**Figure 3.** Pumping speed for $D_2$ *vs* radius of inlet reservoir (dome)

**Figure 4**. Pumping speeds for single-chamber 3-stage pump

**Figure 5**. Indication of where each gas is trapped.

**Figure 6**. Percentage of stage 2 adsorbate that is helium. Sticking coefficient is 0.01, and incoming gas is 2% helium.

**Figure 7.** Percentage of the gas trapped on stage 3 that is helium, different stage 2 sticking coefficients

**Figure 8**. Single-chamber 3-stage pump with 10 sets of panels. The white lines are streamlines.

**Figure 9**. Gas system with 3-stage pump. The number of atoms of each species being trapped on the cryopump are shown, per tritium atom injected by the fuelling system.

**Figure 10**. Illustration of bulk gas flow from regenerating stage 3.

**Supplementary Data Captions**

**DS2VD.DAT**  The DS2V input file for a typical cryopump simulation.

**DS2VD.TXT**  The information from the DS2V input file for a typical cryopump simulation, in human-readable form, as output by the code.